\documentclass[prb,showpacs,twocolumn,superscriptaddress]{revtex4-1}

\usepackage{amsmath}
\usepackage{amsfonts}
\usepackage{amssymb}
\usepackage{graphics}
\usepackage{graphicx}
\usepackage{bm}

\begin{document}

\title{Fast optical control of a coded qubit in a triple quantum dot}

\author{Diego S. Acosta Coden$^1$, Sergio S. Gomez$^1$, Rodolfo H. Romero$^1$, Omar Osenda$^2$, Alejandro Ferr\'on$^1$}

\address{$^1$ Instituto de Modelado e Innovaci\'on Tecnol\'ogica, Universidad Nacional del Nordeste, CONICET, Facultad de Ciencias Exactas y Naturales y Agrimensura, Avenida Libertad 5400, W3404AAS Corrientes, Argentina.}

\address{$^2$ Instituto de F\'{\i}sica Enrique Gaviola, Universidad Nacional de C\'ordoba, CONICET, Facultad de Matem\'atica, Astronom\'{\i}a, F\'{\i}sica y Computaci\'on, Av. Medina Allende s/n , Ciudad Universitaria, CP:X5000HUA C\'ordoba, Argentina}

\begin{abstract}
In this work, we study strategies for the optical control, within the dipole approximation, of a qubit encoded in the three-electron states of a triple quantum dot. 
The system is described by effective confining potentials, and its electronic structure by the configuration interaction method. 
Optimal control theory (OCT) was applied to design low-fluence time-dependent electric fields controlling the qubit in times shorter than a nanosecond. 
The resulting pulses produce transitions between the qubit states for experimentally available field amplitudes with high fidelity.
Their frequency spectra are related to transitions to some lower-lying excited states and a simplified pulse based on those sequential transitions is presented. The limitations of an extended Hubbard description for the type of strategy analyzed here are also discussed.

\end{abstract}
\maketitle
\section{Introduction}
The multiple requirements that a physical system must fulfill to be considered
a worthy qubit have been described extensively. Those candidates 
based on physical systems whose fabrication and control techniques have been driven by industrial applications are a sensible choice. The 
semiconductor physics has been exhaustively developed to provide better materials and devices to the electronic industry.  Semiconductor nano-devices can be constructed
with an ever increasing variety of materials, sizes and shapes\cite{QD1,QD2,QD3}. Besides,  
the number of electrons confined in the device can be chosen precisely. All these 
characteristics result in systems with an scalability and tunability that is lacking in other implementations \cite{Loss1998,DiVincenzo2005,Petta2005,Brunner2011}, in particular
the parameters that define the optical and electrical properties can be 
tailored quite precisely \cite{QD1,QD2,QD3}. 

Physical properties of semiconductor quantum dots with few electrons 
confined in it can be selected to meet and satisfy different criteria\cite{QD1,QD2,QD3,o7,o8}. The controllabilty, {\em i.e.} the ability to change the quantum state of the system in a reliable and repetitive way with a
very high fidelity is most stringent. 
In recent years there has been an increasing interest in controlling
quantum phenomena in molecular systems and nanodevices 
\cite{oct1,oct2,oct3,mur,fosjap,acosta1,acosta3,Rasanen07,Rasanen2008}.
 The coherent quantum control of electrons in quantum dots exposed to
electromagnetic radiation is of great interest, not only in quantum information processing but in
many technological
applications as charge transport devices \cite{oct7,oct8,oct9}. Several studies in quantum control of
double quantum dots (DQDs) has been performed using
gate voltages and optimized laser pulses \cite{oct10,oct11,oct12}.
In addition, the number of quantum control experiments is rapidly
rising through the improvement of laser pulse shaping and closed-loop
learning techniques\cite{oct10,oct11,oct12}.

A qubit state in a few electron quantum dot can be encoded using the electron charge or, as in the case of study of this work, the electron spin. There are a lot of different ways in order to implement a many electron spin qubit. Throughout this work we will analyze the spin manipulation of a three electron 
spin qubit. There exist different ways of implementing a three-spin qubit 
\cite{revbuk}. Among the three electron quantum dot devices for quantum 
information processing we can highlight the exchange-only 
qubit \cite{buk2}, the spin-charge qubit \cite{buk7}, the hybrid qubit 
\cite{buk8,buk9,buk11}, the resonant exchange qubit \cite{buk5,buk6}, and 
the always-on exchange-only qubit \cite{buk16}. These three electron nanodevices
can be implemented confining three electrons in either a single quantum
dot, double quantum dot (DQD), or triple quantum dot (TQD) depending on the 
objective. 

So far, the works dealing with coded qubits \cite{o9} based on the spin states of three electrons confined in a triple quantum dot have investigated the switching between the two basis states forcing the system with potential gates \cite{o10}. The basis states of the coded qubit are two eigenstates with total intrinsic angular momentum $S_z=1/2$. Switching times on the order of tens or hundreds of nanoseconds are commonly achieved and, since the decoherence mechanisms have characteristics time scales of microseconds, the number of controlled operations that could be performed is relatively low. Regrettably, with gated operations it is not possible to use frequencies beyond 
hundreds of Gigahertz so sub nanoseconds operations, and consequently larger number of 
controlled operations, are beyond the reach of gated systems \cite{o11}. More recently a similar proposal that uses a time-dependent electric field to drive a molecular magnet promises operation times on the picoseconds order \cite{o12}. So, it seems reasonable to explore the manipulation of the coded qubit using as control a time-dependent electric
field with a frequency on the Terahertz range to achieve operation times in the order of the picoseconds. Of course, using an electromagnetic field introduce a set of difficulties to overcome. As we will show, there is no need to address the electrons individually, it is just enough that the field strength remains constant along the spatial region that contains the TQD, which will have a characteristic length around $100$ nanometers. The dynamical behaviour and control of coded qubits based on spin states of three-electron TQD have been analyzed theoretically rather thoroughly, using  Hubbard Hamiltonian approximations, or effective Heisenberg Hamiltonians \cite{o9,o13,o14}. These approaches were justified since, for a broad range of parameters, the qubit basis states are far away from the other three-electron states. Besides, for the driving frequencies and forcing amplitudes accessible to gate technology, the system shows well-defined Rabi oscillations. Since we will consider frequencies quite different to those applied by metallic gates, neither  a simple sinusoidal driving would result in fast Rabi oscillations nor a model that fits the two lowest eigen-energies values would predict accurately the dynamical behaviour. Is in scenarios like the one described where Optimal Control Theory provides the means to design sequences of composite control pulses.

The aim of this work is to present a detailed analysis of the optical control of three electrons in a two-dimensional triple quantum dot by means of Optimal Control Theory. The paper is organized as follows. In Sec. 2 we introduce the model for the two-dimensional three-electron triple quantum dot (2A) and briefly describe the method used to calculate its electronic structure (2B). In Sec. 2C, we describe optimal control equations for the two-dimensional three-electron quantum dot device. In Sec. 3A we analize the electronic structure of the device using different approaches and in Sec 3B we study the operation of the coded qubit by means of OCT. Finally, in Sec. 4, we discuss our most relevant results.

\section{Model and Methods}
\subsection{Model of the triple quantum dot}
\begin{figure}
\begin{center}
$\begin{array}{c}
\includegraphics[width=8cm]{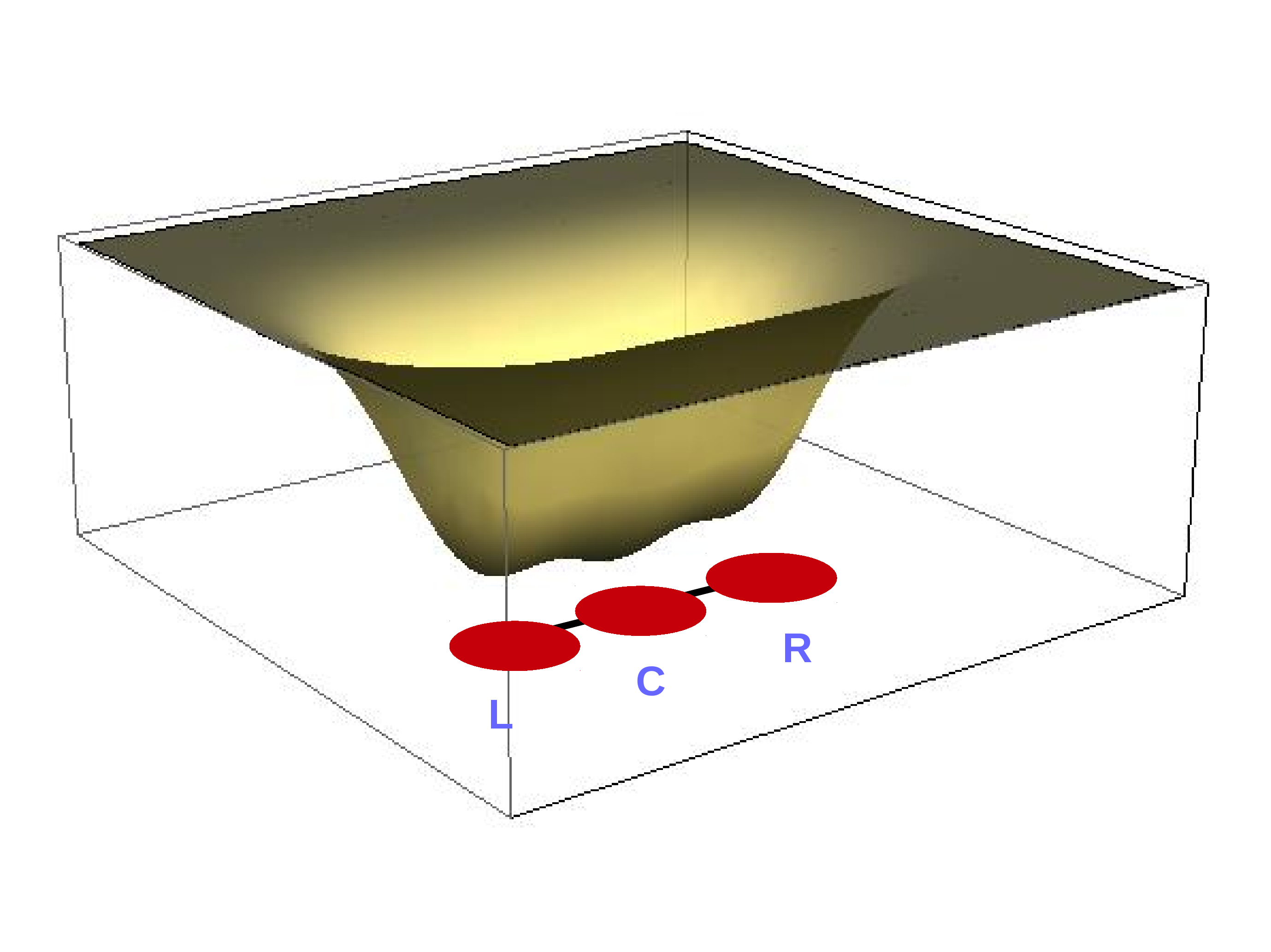} 
\end{array}$
\caption{\label{TQD scheme} Scheme of the linear arrangement of three quantum dots named as left (L), center (C) and right (R). The red circles in the horizontal $x$-$y$ plane represent the confinement regions of charge of the two-dimensional electron gas formed in the interface of a typical heterostructure. The confining  potential used in this work, consisting in a superposition of a Gaussian well for each one, is depicted above the plane. }
\end{center}
\end{figure}

We consider a linear triple quantum dot (LTQD) with three electrons. The $x$ axis is chosen along the LTQD such that the dots are centered at positions ${\bf R}_n=(x_n,0)$ ($n = 1, 2, 3$), as shown in the Fig. \ref{TQD scheme}. 
The model Hamiltonian is $H=H_0+V(t)$, where
\begin{equation}
H_0 = \sum_{i=1}^3 h({\bf r}_i) + \sum_{i, j=1}^3 V_C(|{\bf r}_i-{\bf r}_j|) ,
\end{equation}
is the time independent Hamiltonian that gives rise to the level structure of the LTQD.
\begin{equation}
h({\bf r}) = -\frac{\hbar^2}{2m^*}\nabla^2+\sum_{i={\rm L, C, R}} W_i({\bf r})
\label{hamiltonian 1-electron}
\end{equation}
contains the two-dimensional  kinetic energy and the confinement potential energy $W_i({\bf r})$, which in this work is modeled by a Gaussian well 
\begin{equation}
W_i({\bf r}) = -W_i \exp\left( -\frac{|{\bf r}-{\bf R}_i|^2}{2a^2}\right), 
\end{equation}
of depth $W_i$ and range $a$. In quantum dots electrostatically produced, their size and separation can be both controlled by tunable gate voltages through metallic electrodes deposited on the heterostructure interface. 
In the model, setting the depths $W_i$ ($i=$ L, C, R) of each dot simulates the experimental tuning by the electrostatic plunger gates $V_1$, $V_2$ and $V_3$ respectively. Usually, the regime of work of the device is characterized by two detuning parameters, $\varepsilon=V_1-V_3$ and $\varepsilon_m=V_2-(V_1+V_3)/2$. 
$V_C(|{\bf r}_i-{\bf r}_j|)=   e^2/4\pi\epsilon\epsilon_0 r_{ij}$ is the electron-electron Coulomb repulsion; $e$ and $m^*$ are the effective charge and mass of the electrons in the heterostructure. 

Finally
\begin{equation}
V(t) = -e{\bf F}(t){\bf \cdot}\sum_i {\bf  r}_i = -eF(t)\sum_i x_i
\label{time-dep V}
\end{equation}
represents the interaction with an external uniform time-dependent electric field ${\bf F}(t)$ along the LTQD and is the Hamiltonian used for controlling the evolution of the system. 

The numerical results presented in this work refers to those corresponding to the parameters of GaAs: effective mass $m^*= 0.067 m_e$, effective dielectric constant $\epsilon =13.1$, Bohr radius $a^*_B = 10$ nm and effective atomic unit of energy 1 Hartree$^*= 10.6$ meV. The confining potentials at zero detuning ($\varepsilon=\varepsilon_m=0$) are $W_{\rm L}=W_{\rm R}=53$ meV, $W_{\rm C}=51$ meV for the three  wells to have the same depth, and $a=16.3$ nm, with the central dot separated 35 nm from the left and right ones.
\subsection{Electronic structure}
We use the full configuration interaction (CI) method to calculate the stationary three-electron states  of the LTQD, in the absence of  the time-dependent field and in any subspace with defined $S_z$, where $S_z$ is the $z$ component of the total spin ${\bf S}={\bf S}_1+{\bf S}_2+{\bf S}_3$. We mainly focus in $S_z=1/2$ since we know that for three-electron system this subspace contains both the ground state and the first excited state with this projection of total spin. Also, because we are interested in the electrical manipulation of the system, the states with different $S_z$ will not play a role in our study and we will not consider them.  

We change to the second-quantization representation of $H_0$ 
\begin{equation}
H_0 = \sum_{ij,\sigma}\varepsilon_{i}c_{i\sigma}^\dag c_{i\sigma} +\frac{1}{2} \sum_{\substack{i,j,k,l \\ \sigma\sigma'}} V_{ijkl}c_{i\sigma}^\dag c_{j\sigma'}^\dag c_{k\sigma'} c_{l\sigma},
\label{H_0 second-quantized}
\end{equation}
where the second-quantization creation ($c_{i\sigma}^\dag$) and annihilation ($c_{i\sigma}$) operators are such that $|\varphi_{i\sigma}\rangle=c_{i\sigma}^\dag|vac\rangle$, with the one-electron wave functions $\varphi_{i\sigma}$ being solutions of the eigenvalue problem $h\varphi_{i\sigma} = \varepsilon_i \varphi_{i\sigma}$, $h$ is the one-particle Hamiltonian [eq.  (\ref{hamiltonian 1-electron})] and $\sigma$ labels the eigenstates of up or down ($\pm 1/2$) projections of $\sigma_z$. Furthermore,
\begin{equation}
 V_{ijkl} = \int \varphi_{i}^{*}({\bf r}_1) \varphi_{j}^{*}({\bf r}_2)\frac{e^2}{4\pi\epsilon\epsilon_0 r_{12}} \varphi_{k}({\bf r}_2)\varphi_{l}({\bf r}_1) d{\bf r}_1 d{\bf r}_2
\end{equation}
are the two-particle matrix elements of the Coulomb interaction.

The three-particle eigenstates $|\Psi_n\rangle$ of the subspace $S_z=1/2$ of $H_0$ are obtained by exact diagonalization of $H_0$ in a basis of configurations of antisymetrized products $|\Phi_{ijk}\rangle\equiv c_{i\uparrow}^\dagger c_{j\uparrow}^\dagger c_{k\downarrow}^\dagger |{\rm vac}\rangle $ of two spin-up and one spin-down orbitals of $h$  
\begin{equation}
|\Psi_n\rangle = \sum_{ijk} a_{ijk}^{(n)} |\Phi_{ijk}\rangle .
\end{equation}
The time-dependent Hamiltonian $V(t)$, eq. (\ref{time-dep V}),
\begin{equation}
V(t) = \sum_{ij,\sigma} V_{ij}c_{i\sigma}^\dag c_{j\sigma}
       = - F(t) \sum_{ij,\sigma} \mu_{ij}c_{i\sigma}^\dag c_{j\sigma},
\end{equation}
with $\mu_{ij}=\langle \varphi_i |e x| \varphi_j\rangle$, induces transitions between the different levels of $H_0$ although, since $H$ is spin-independent, there are no transitions between subspaces of different total spin $S$.\\

The implementation and the accuracy of the CI method depends on the number of one-electron wave functions used in the calculations. Results present along this work were performed using 9 orbitals obtained by direct solution of Eq. (\ref{hamiltonian 1-electron}) \cite{acosta3,jpb}. Nevertheless, the large number of orbital configurations entering in an accurate calculation spoils a simple interpretation of the three-electron LTQD states in terms of configurations of dot occupations.

On the other hand, it has been shown \cite{Yang2011,Burkard99} that many quantum dot properties can be described in simple terms with a properly parameterized extended Hubbard approximation to the Hamiltonian (\ref{H_0 second-quantized}), namely,
\begin{eqnarray} 
H_{\rm Hub} &=& \sum_{\langle i,j \rangle, \sigma=\uparrow,\downarrow} \tau_{ij} (d^\dag_{i\sigma} d_{j\sigma} + {\rm h.c.}) \\    \nonumber 
&& +  \sum_{i=1}^N \left[\frac{1}{2}(U-U_C) n_i(n_i-1) + V_i n_i \right] \\ \nonumber
&&+ \sum_{\langle i,j \rangle} U_C n_i n_j,
\end{eqnarray}
where $\tau_{ij}$ are nearest-neighbor hopping parameters, $U$ and $U_C$ are respectively the on-site and  intersite repulsion, and $V_i$ is the gate electrostatic potential at dot $i$; $d^\dag_{i\sigma}$ ($d_{i\sigma}$) creates (annihilates) an electron localized  at dot $i$, and $n_{i}=\sum_{\sigma=\uparrow,\downarrow} d^\dag_{i\sigma} d_{i\sigma}$ is the occupation of dot $i$. 

The effect of the time-dependent electric field is accounted for by the matrix elements of the one-electron operator 
\begin{equation}
V(t)=-\mu F(t)=-F(t)\sum_{ij,\sigma=\uparrow,\downarrow} \mu_{ij}c_{i\sigma}^\dag c_{j\sigma}
\end{equation}
between the $H_{\rm Hub}$ eigenstates, where the explicit form of $\mu$ in the three-electron configuration basis set is shown in the  \ref{Apendix-hubbard}  

For the  Hubbard model calculations, we shall assume a negligible overlap between one-electron states localized at the dots, so that, the dipole moment operator becomes diagonal and nonvanishing for double occupied configurations only, i.e., $\mu={\rm diag}(0,0,\lambda,-\lambda,\lambda,-\lambda)$, with $\lambda=\langle\varphi_R|e x|\varphi_R\rangle$.
For the sake of comparison to CI calculations with confinement potentials, we chose hopping matrix elements $\tau_{ij}=\tau=1.0$ meV, Coulomb on-site repulsion $U=6.4$ meV and intersite repulsion $U_C=1.6$ meV, so that they become qualitatively comparable to each other.

\subsection{Optimal control of the state dynamics}
For controlling the qubit, encoded in the two lowest LTQD states $\Psi_0$ and $\Psi_1$, we propose to design a time-dependent electric field $F(t)$ able to drive the state of the system $\Psi(t)$ from $\Psi(0)=\Psi_0$ to $\Psi(T)=\Psi_1$, in a given prescribed time $T$ as short as possible. 
We assume the time-dependent electric field $F(t)$ to be uniform and applied along the longitudinal direction $x$ of the LTQD. Then, the state $\Psi(t)$ must satisfy the time-dependent Schr\"odinger equation
\begin{equation}
i\frac{\partial \Psi(t)}{\partial t}=H
\Psi(t) =
[H_0-\hat{\mu}F(t)]\Psi(t). 
\label{TDSE}
\end{equation}
Optimal control theory (OCT) provides a systematic method to find the optimal pulse $F_{\rm OCT}(t)$ that drives the solution of the Schr\"odinger equation (\ref{TDSE}), while maximizing the overlap of the final state to a given target state $\phi_F$. 
We briefly outline here the theory and some technical details of the method as applied in this work \cite{oct16,moct1}.
 
The yield of the control process is defined by the following functional
\begin{equation}
J_1[\Psi]=|\langle\Psi(t)|\phi_F\rangle|^2.
\end{equation}
In order to avoid high energy fields a second functional is introduced
\begin{equation}
J_2[{\bf\varepsilon}]=-\alpha\left[\int_0^T F^2(t)dt\right],
\end{equation}
where the fluence is the time-integrated intensity of the field and $\alpha$  is a time-independent Lagrange multiplier. As we already know, the electronic wavefunction has to satisfy the time-dependent Schr\"odinger equation introducing a third functional
\begin{equation}
J_3[{\bf\varepsilon},\Psi,\chi]=-2Im \int_0^T \left\langle\chi(t)\left|\frac{\partial}{\partial t}-H(t)\right
|\Psi(t)\right\rangle
\end{equation}
where we have introduced the time-dependent Lagrange multiplier $\chi(t)$. Finally, the Lagrange functional has the form $J=J_1+J_2+J_3$. The Variation of this functional with respect to 
$\Psi(t)$, $F(t)$ and $\chi(t)$ allows us to 
obtain the desired control equations \cite{moct1}
\begin{eqnarray}
i\frac{\partial \Psi(t)}{\partial t}&=&H
\Psi(t)\;,\; \Psi(0)=\phi\\
i\frac{\partial \chi(t)}{\partial t}&=&H(t)\chi(t) \\  \chi(T)&=&|\phi_F\rangle\langle\phi_F|\Psi(T)\rangle\\
F(t)&=&-\frac{1}{\alpha}Im\langle\chi(t)|\hat{\mu}|\Psi(t)\rangle
\end{eqnarray}
This set of coupled equations can be solved iteratively. The solution can be obtained using several approaches, for example, using the efficient forward-backward propagation scheme developed in \cite{oct17}. The algorithm starts with propagating $| \phi \rangle$ forward in time, using in the first step a guess of the laser 
field $F^{(0)}(t)$. At the end of this step we obtain the
wavefunction $| \Psi^{(0)}(T) \rangle$, which is used to evaluate
$| \chi^{(0)}(T) \rangle  =| \phi_F\rangle\langle\phi_F|\Psi^{(0)}(T )\rangle$. The algorithm  continues with propagating $|\chi^{(0)}(t) \rangle$ backwards in time. In this step it is clear thet we need
to know both wavefunctions ($\Psi^{(0)}$ and $\chi^{(0)}$) at the same time. In this step we obtain a first version of the optimized pulse $F^{(1)}(t)$. We repeat this operation until the convergence of $J$ is achieved. The values of the  Lagrange multiplier $\alpha$ are chosen in order to achieve the desired fluences. The numerical integration of the forward and backward time evolution was performed using high precision Runge-Kutta algorithms under various initial guesses (e.g., constant, random, monochromatic and a superposition of two harmonic pulses). The resulting optimal pulse and the time evolution under its driving were found to be robust for all these various starting conditions.
\section{Results}
\subsection{Energy spectra and dipole transition moments}

\begin{figure}
\begin{center}
 \includegraphics[width=8cm]{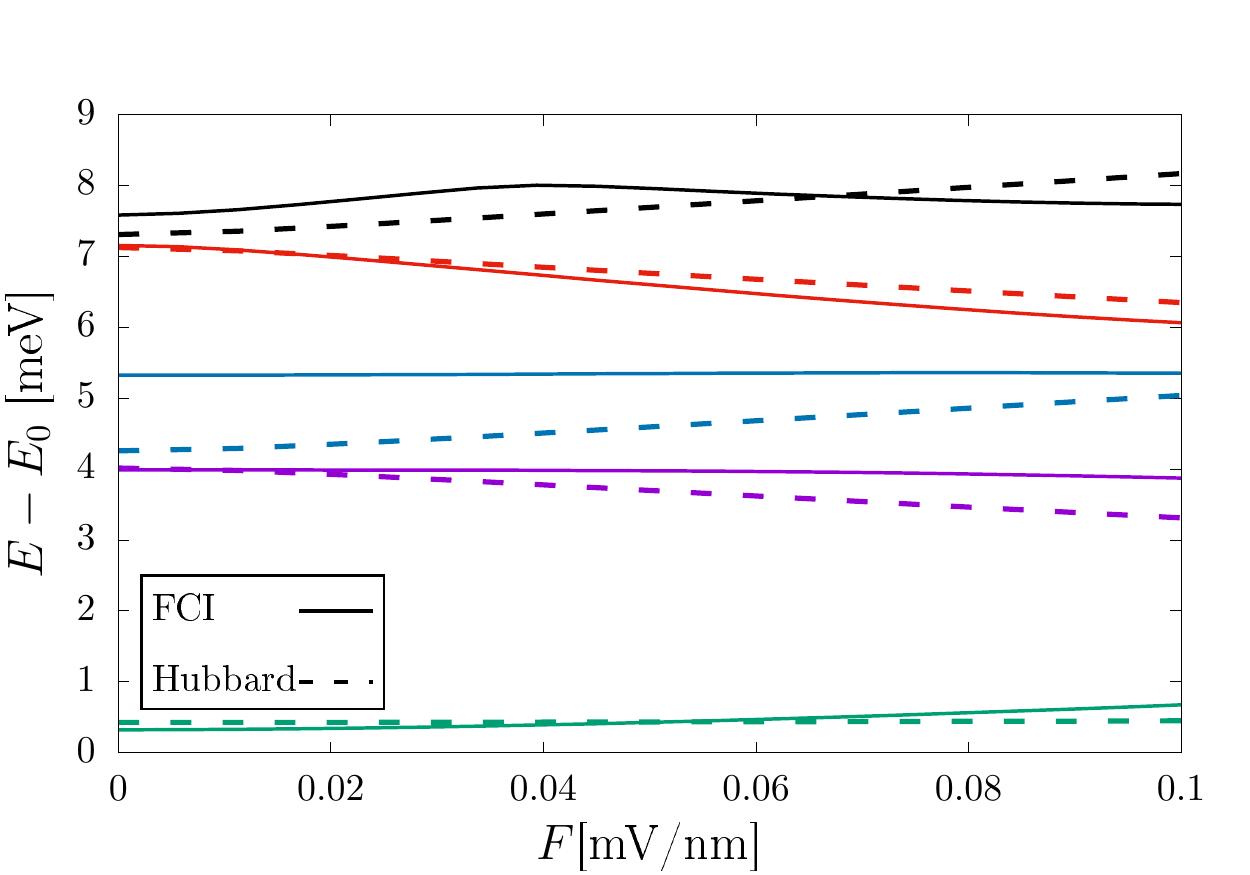} 
\caption{\label{Energy vs Field} Energy gaps $E_n-E_0$ of the TQD levels referred to the ground state as a function of the electric field $F$ calculated with the CI method (solid lines) and the Hubbard model (dashed lines). As discussed in the text, the Hubbard model describes qualitatively well the gaps, with the exception of $n=3$. Altough hubbard predicts a small energy gap $E_3-E_2$ smaller than 0.1 meV, the most accurate result given by FCI calculations is more than 1 meV     }
\end{center}
\end{figure}


The configuration interaction method provides very accurate numerical approximations for both, the eigenvalues and eigenfunctions of systems of interacting electrons, as long as a large enough number of one-electron wave functions are used to perform the calculations. The large number of one electron functions needed to obtain the desired accuracy in the three-electron energies implies in general a large number of configurations, and therefore a representation of the hamiltonian provided by the Hubbard model could allow a more direct and simple interpretation of the relevant configurations on a site  basis set, that is, a basis of orbitals localized around a given Dot. By using this simple model, the symmetry of total spin, and the projection in the z axis, the number of configurations can be reduced to only six in total. This simple model can provide insights of the mechanism involved in the effects of an electric field in the system. Then, in the present section we compare the results of the electronic structure and the response to an electric field obtained with CI calculations and the Hubbard model presented in the previous section, in order to know in which aspects both approximations describe the results with the same features.           

Fig. \ref{Energy vs Field}  shows the energy levels gaps with respect to the ground state of the LTQD calculated with the CI method and the Hubbard model, as a function of the magnitude of an applied external uniform electric field $F$. 
At zero field, the energy levels are roughly grouped as $(E_0,E_1)$, $(E_2,E_3)$ and $(E_4,E_5)$ with relatively small gaps between them. For the CI calculations we obtain  $E_1-E_0 \simeq 0.23 $meV , $E_3-E_2 \simeq 1$ meV and $E_5-E_4\simeq 0.64 $ meV.

The Hubbard model at zero detuning provides a simple picture for this structure. The two lowest lying levels are single occupied; the second and the highest pairs are double occupied dots configurations of higher energies, $U-2U_C$  and $U$, respectively, with small gaps between the levels of each pair $\Delta E\sim \tau^2/U$. As shown in the figure, the agreement of the energy gaps is fairly good, with the main qualitative difference between the CI and Hubbard being the gap of the pair $(E_2,E_3)$, which is underestimated within the approximate model. This discrepancy has consequence on the dynamics of the qubit as we shall discuss in the section below.

The Hubbard model provides insight on the configurations forming the states from CI calculations. 
At zero external field and vanishing detuning $V_2=0$, the Hubbard eigenstates  are approximately
\begin{eqnarray}
|0^{1/2}\rangle &\approx&|0\rangle \\
|1^{1/2}\rangle &\approx& |1\rangle \\
|\Psi_2\rangle &\approx& \frac{1}{\sqrt{2}}(|2\rangle +|3\rangle)\\
|\Psi_3\rangle &\approx& \frac{1}{\sqrt{2}}(|2\rangle - |3\rangle) \\
|\Psi_4\rangle &\approx& \frac{1}{\sqrt{2}}(|4\rangle +|5\rangle)\\
|\Psi_5\rangle &\approx& \frac{1}{\sqrt{2}}(|4\rangle -|5\rangle), 
\end{eqnarray}
where the logic states $|0^{1/2}\rangle$ and $|1^{1/2}\rangle$ are configurations of single occupied dots, while $|\Psi_n\rangle$ ($n=2,\ldots, 5$) involve configurations with double occupied dots. 

Application of an external electric field $F$ produces a mix of single and double configurations, such that $|\Psi_i(F)\rangle = \sum_n c_{in}(F) |n\rangle$, 
where the coefficients $c_{in}$ stand for contributions of single and double occupied configurations, having magnitudes $c_{in}\sim{\cal O}(\tau/U)\ll 1$. Since any one-particle operator involves the hopping of a single electron from a dot to its nearest neighbor, it cannot connect single occupied states (nor double occupied states) to each other. Thus, this small mixing between single and doubly occupied states allows for dipole transitions among the energy levels.

\begin{figure*}
\begin{center}
$\begin{array}{cc}
\includegraphics[width=8cm]{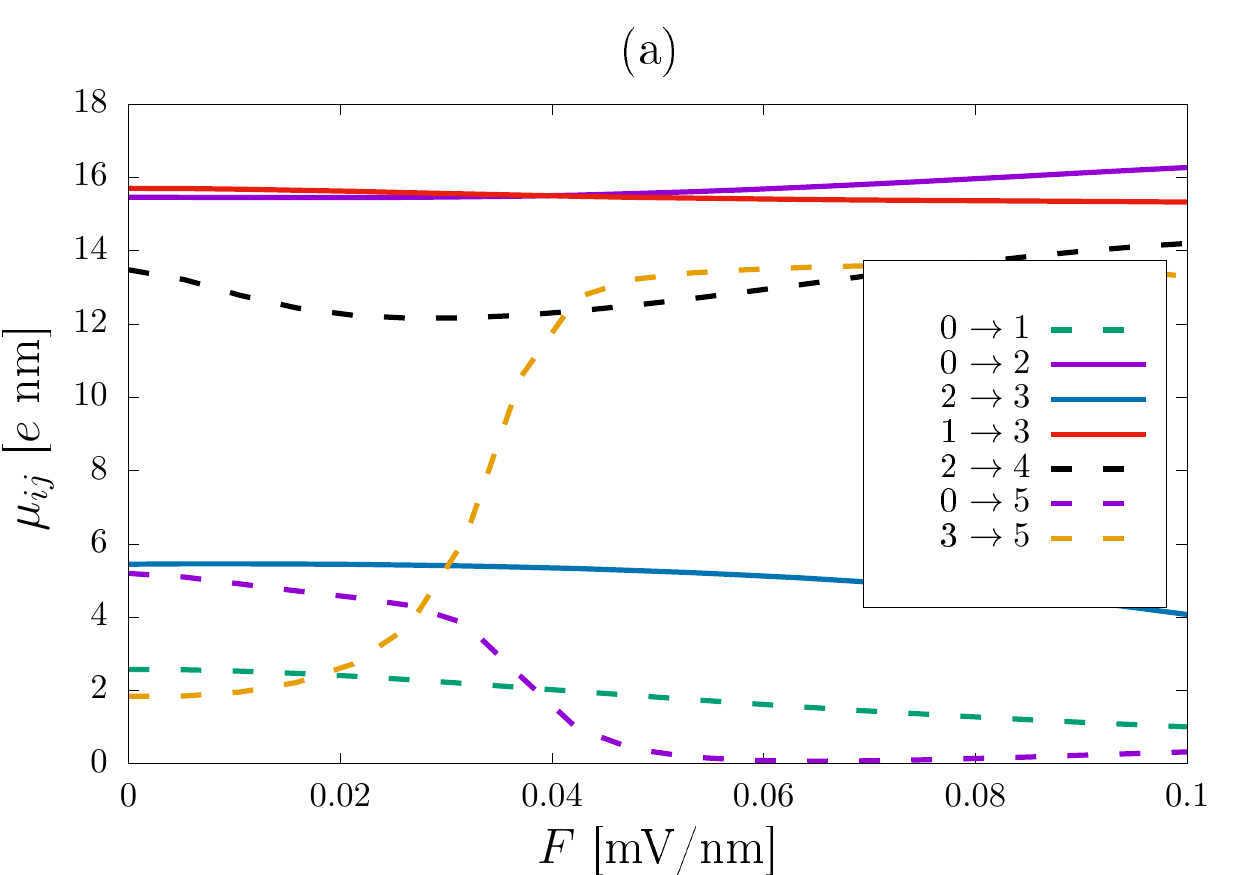} &
\includegraphics[width=8cm]{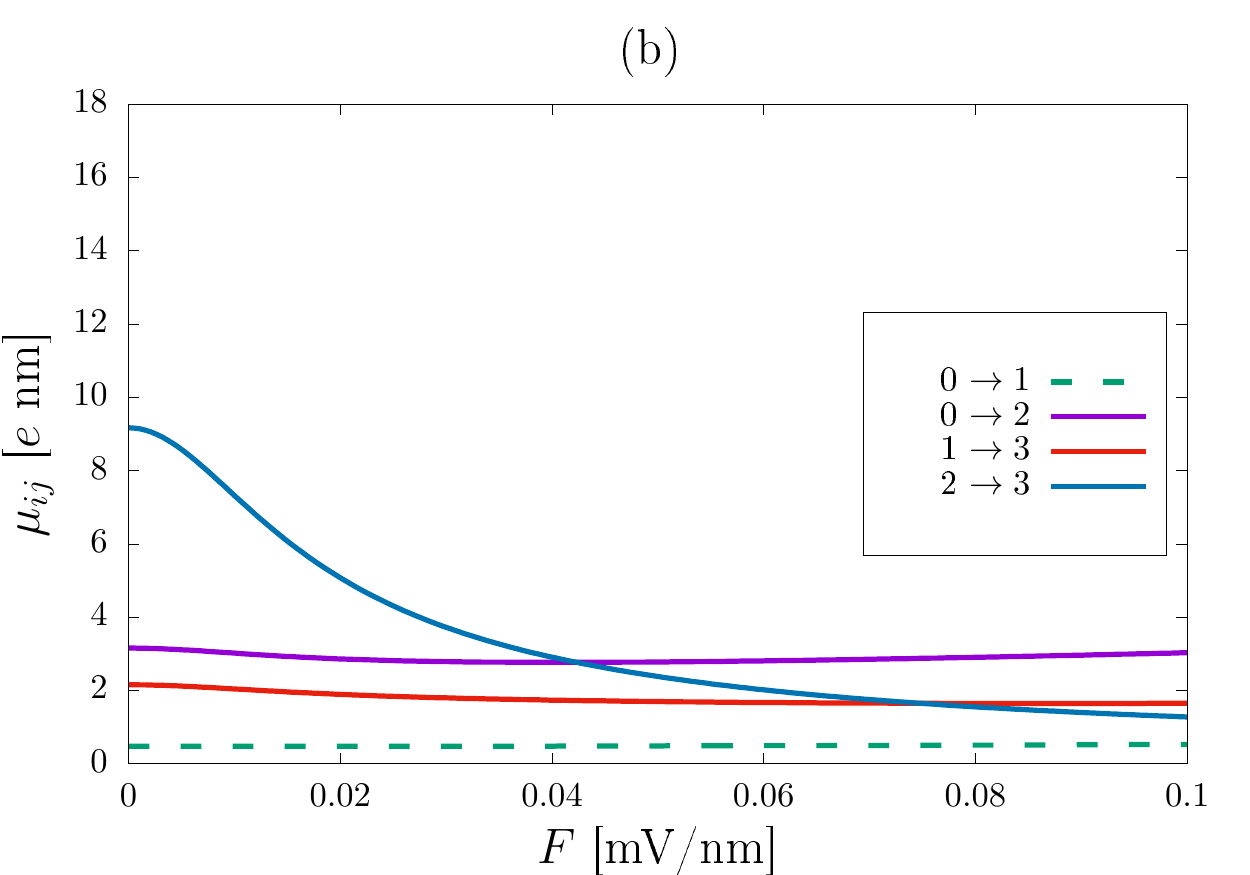} 
\end{array}$
\caption{\label{dipole vs F}
Relevant transition dipole moments between the eigenstates of the (a) CI  and (b) Hubbard  Hamiltonians as a function of the electric field $F$. The Hubbard parameters are $\tau=1$ meV, $U=6.4$ meV and $U_C=1.6$ meV . See the text for details. In the figure the unit for the dipole moment is given by  $e$ nm , where $e$ is  the electron charge.}
\end{center}
\end{figure*}

For the transfer of interest, $0\rightarrow 1$, the particular evolution followed by the system driven by a laser time-dependent electric field, shall depend on the magnitude of the transition moments $\mu_{ij}$ coupling the lowest-lying states. 
Figs. \ref{dipole vs F}(a) and \ref{dipole vs F}(b) show the variation of the dipole moment matrix elements $\mu_{ij}$ with the electric field $F$, obtained from the full CI method and Hubbard model calculations, respectively. 
The transitions moments $\mu_{02}$, $\mu_{23}$ and $\mu_{31}$, are shown in solid lines, while other transitions are shown with dashed lines, for comparison of two distinct mechanisms, namely, a {\em direct} transfer $0\rightarrow 1$ and an {\em indirect} one $0\rightarrow 2 \rightarrow 3\rightarrow 1$. 

A direct comparison of the CI and Hubbard results (Fig. \ref{dipole vs F}) shows that, for the direct transition, they differ approximately by an order of magnitude ($\mu_{01}^{\rm Hubb}\ll \mu_{01}^{\rm CI}$) with $\mu_{01}^{\rm Hubb}$ being almost constant while $\mu_{01}^{\rm CI}$ decreasing slightly with increasing $F$. 
Interstingly, the CI calculation gives a nearly constant $\mu_{02} \simeq \mu_{31}$ for a wide range of fields, that is approximately satisfied by the Hubbard results. Nevertheless, also in this case, the $\mu^{\rm CI}$ becomes much larger than $\mu^{\rm Hubb}$.
The most remarkable difference between both descriptions is the transition $\mu_{23}$, for which the Hubbard calculation overestimates the CI value and becomes strongly dependent on the field.

These results, together with the already mentioned difference in the gap $E_3-E_2$, illustrate the limitations of the Hubbard model to describe the transitions when doubly occupied configurations are relevant. Therefore, all dynamical processes discussed below, were only performed with magnitudes calculated with the CI method.

Finally, other paths differing from the direct and indirect ones, such as $0 \rightarrow 2 \rightarrow 1$ or $0 \rightarrow 3 \rightarrow 1$, are ruled out due to a low or vanishing dipole moment between a given pair or states. On the other hand, transitions involving higher excited states, like $0 \rightarrow 5 \rightarrow 3 \rightarrow 1$ are not found in the OCT solution to be discussed below, possibly because of the requirement of low fluence of the pulse.

\subsection{Qubit dynamics}
	
The transition dipole moments calculated with the CI wave functions were used to optimally design fast low-fluence pulses to control the qubit levels $\{|0^{1/2}\rangle, |1^{1/2}\rangle\}$. 

Due to the imposed joint constrains on the pulse length and the low fluence, the resulting pulse is not always able to accomplish a complete population transfer to the target state. Pulses of very short lengths ($T\lesssim 10$ ps) and low fluence, can transfer the initial state with $\lesssim 0.9$ fidelity, which is lesser than the required accuracy for quantum information applications.  

When the amplitud of the applied pulse is near the value $F=\pi/\mu_{01}T$ (i.e., small values of $\alpha$), the OCT method favours monochromatic pulses driving a direct Rabi transition with frequency  $\omega_{01} \simeq$ 540 GHz. By choosing larger values of $\alpha$, the amplitude of the resulting pulse becomes smaller, and Rabi transitions are not longer possible for the fixed high yield. Nevertheless, in those cases, there are other solutions, involving transitions to excited states, that can perform the transfer process within the imposed constraints. We shall focus here on these low-amplitude fast pulses able to drive the state with high-yields, although at the expense of using transitions to the low lying excited states, external to the logical computational subspace.     

All the calculations presented in the following are in a range of pulse length and fluence for which the tailored pulses reach 99\% (or higher) of population transfer. 
\begin{figure*}
\begin{center}
$\begin{array}{cc}
(a) & (b) \\
\includegraphics[width=6cm]{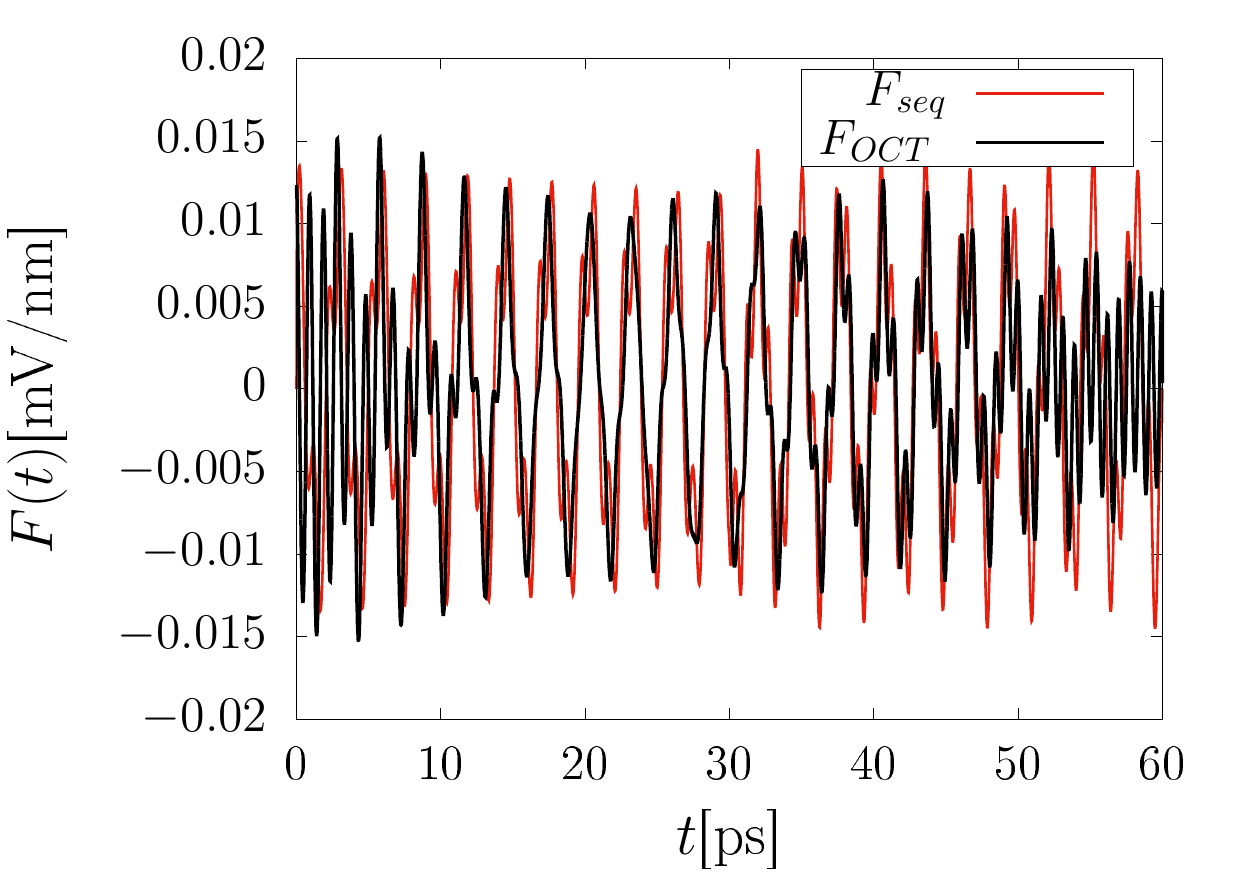} &
\includegraphics[width=5.5cm]{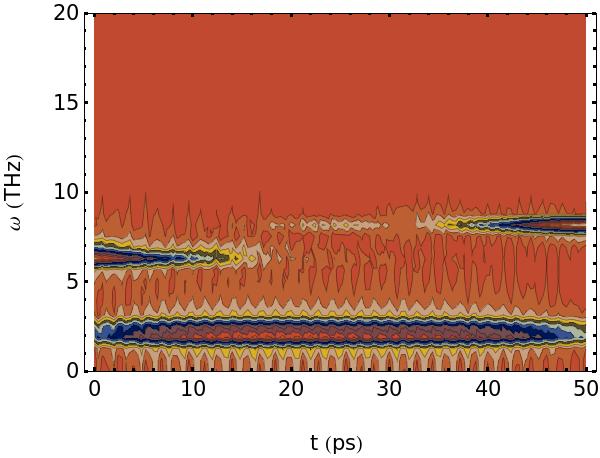} \\
(c) & (d) \\ 
\includegraphics[width=6cm]{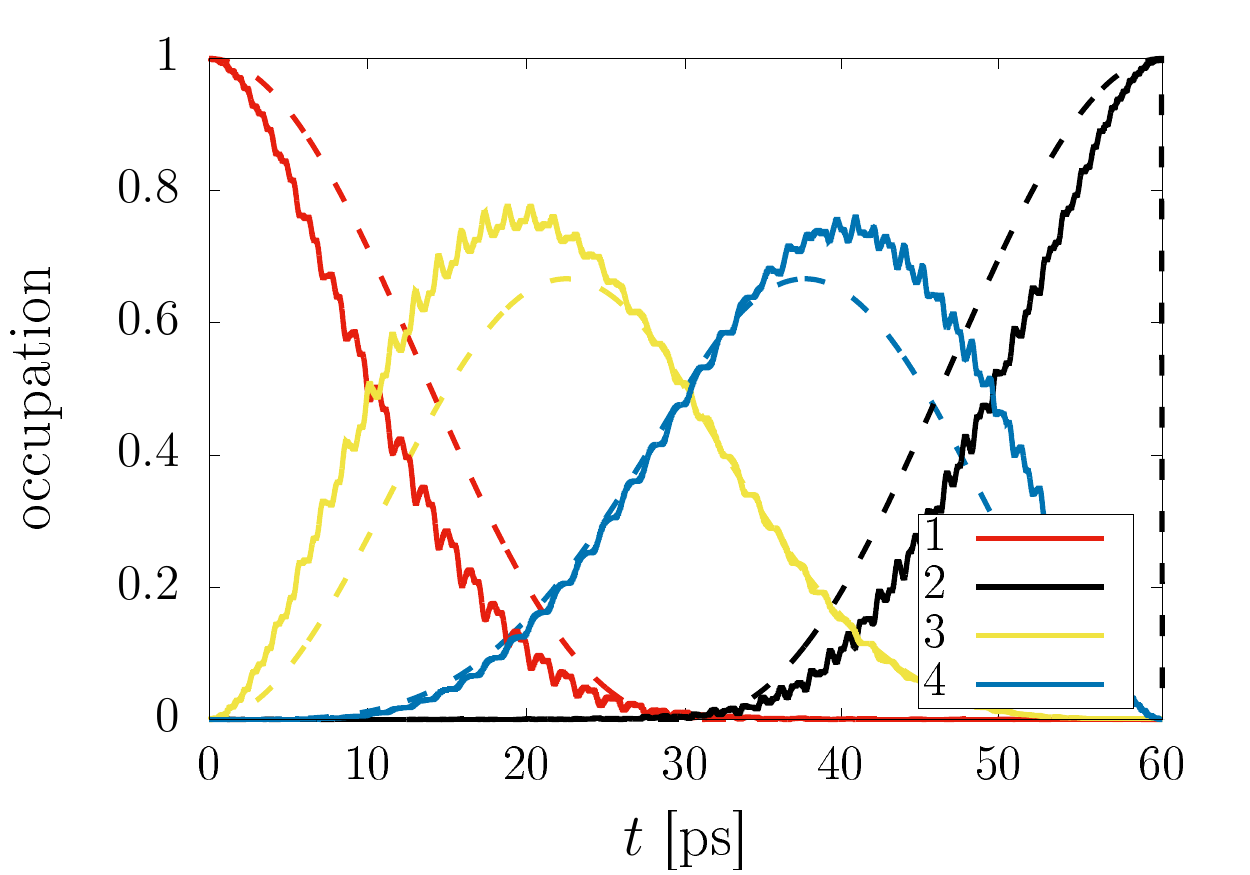} &
\includegraphics[width=6cm]{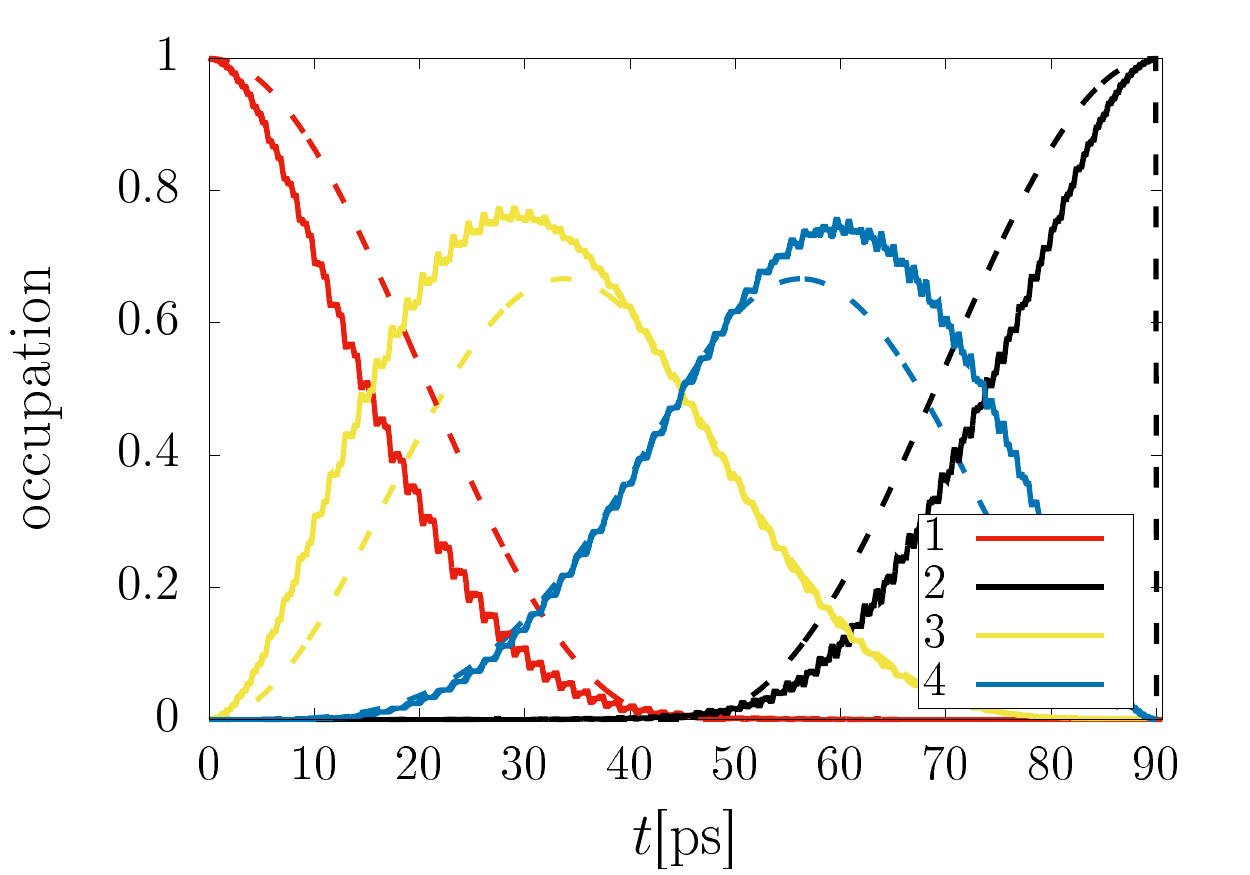}
\end{array}$
\caption{\label{OCT} (a) Optimally (OCT) designed and three-level model pulses for $0 \rightarrow 1$ population transfer in $T=60$ ps. (b) Map of Fourier transform  of the OCT pulses $F(t)$ calculated within a time window ($t, t+\Delta t$) as a function of $t$ and $\omega$, with $\Delta t=10$ ps. Lower panels: Time evolution of the population of the four lowest states resulting from driving the system with the pulses of $T=60$ (c) and $T=90$ ps (d).}
\end{center}
\end{figure*}

A typical pulse $F_{\rm OCT}(t)$ obtained from the OCT optimization, for a pulse length of  $T=60$ ps, is shown in Fig. \ref{OCT}(a) as a black solid line together with the sequential pulse $F_{\rm seq}(t)$, eq. (\ref{pulse seq}), to be discussed below (red thin line).

A Fourier analysis of $F_{\rm OCT}(t)$ shows strong contributions from three frequencies:  $\omega_{20}=6.73$ THz,  $\omega_{32}=2.36$ Tz and $\omega_{31}=8.54$ Tz, corresponding to the energy gaps of the TQD: $\hbar\omega_{20}$=4.18 meV and $\hbar\omega_{32}$=1.47 meV and $\hbar\omega_{31}$=5.31 meV. 
These results suggest that the dynamics can be viewed as transitions along the paths connecting the initial and target states and having the largest transition dipole moments, i.e., $\Psi_0\rightarrow\Psi_2\rightarrow\Psi_3\rightarrow\Psi_1$.

The time dependence of the pulse is described in more detail in Fig. \ref{OCT}(b). It shows a map of the Fourier transform of the slice of $F_{OCT}(t)$ in a time window $(t,t+\Delta T)$, for $\Delta T=10$ ps, as a function of the frequency $\omega$ and $t$ $(0\le t\le 50)$ ps. It can be seen that while the lowest frequency $\omega_{32}$ is present all along the pulse length $T$, $\omega_{20}$ and $\omega_{31}$ are present, roughly, during the first and second halves, respectively. 
The dynamics of the state occupations, Fig. \ref{OCT}(c), confirms this picture. The pulse produces a transfer of population exclusively with the state 2 during the first 10 ps; then, it keeps transferring population to 2, but also an exchange $2\rightarrow 3$ occurs until the state 0 becomes completely depleted at nearly $T/2$. In the second half $(T/2,T)$, a symmetrical process occurs with the occupation of the target state 1 increasing by exchanging population with state 3 while the interchange $2\rightarrow 3$ holds.
Fig. \ref{OCT}(d) shows that similar processes occur for a longer pulse with $T=90$ ps.\\

From the discussion above, a simpler two-frequencies sequential pulse is proposed for $0\le t\le T$:
\begin{equation}
    F_{\rm seq}(t)=F_1(t)\theta(T/2-t)+F_2(t)\theta(t-T/2), 
    \label{pulse seq}
\end{equation}
where $\theta(t)$ is the step function, and
\begin{eqnarray}
F_1(t) &=& A {\rm cos}(\omega_{20} t) + B {\rm cos}(\omega_{32} t), \\
F_2(t) &=& A {\rm cos}(\omega_{31} t) + B {\rm cos}(\omega_{32} t),
\end{eqnarray}
where the assumptions that the same amplitude $A$ for the transitions $0\rightarrow 2$ and $3\rightarrow 1$ in the ranges $(0,T/2)$ and $(T/2,T)$, respectively, and a constant amplitude $B$ for $2\rightarrow 3$ along the whole range $T$ are made. 

The analytic solution of the time-dependent Schr\"odinger equation for this pulse, within the rotating wave approximation (RWA), is given in   \ref{ApendixA}. The resulting pulse for $T=60$ ps, shown in Fig. \ref{OCT}(a) with red thin line, resembles to a large extent the optimal pulse. Also the time variations of the occupations when this pulse is applied, dashed curves in Figs. \ref{OCT}(c) and \ref{OCT}(d), represent a good approximation to those obtained with optimal pulse.

In the usual description of three level systems, the application of $F_{\rm seq}(t)$ can be considered as two sequential three-level processes, namely, a ladder-transition ($0\rightarrow 2 \rightarrow 3$) with $F_1(t)$, followed by a $\Lambda$-process ($2\rightarrow 3\rightarrow 1$) with $F_2(t)$.

\begin{figure}
$\begin{array}{c}
\includegraphics[width=7.25cm]{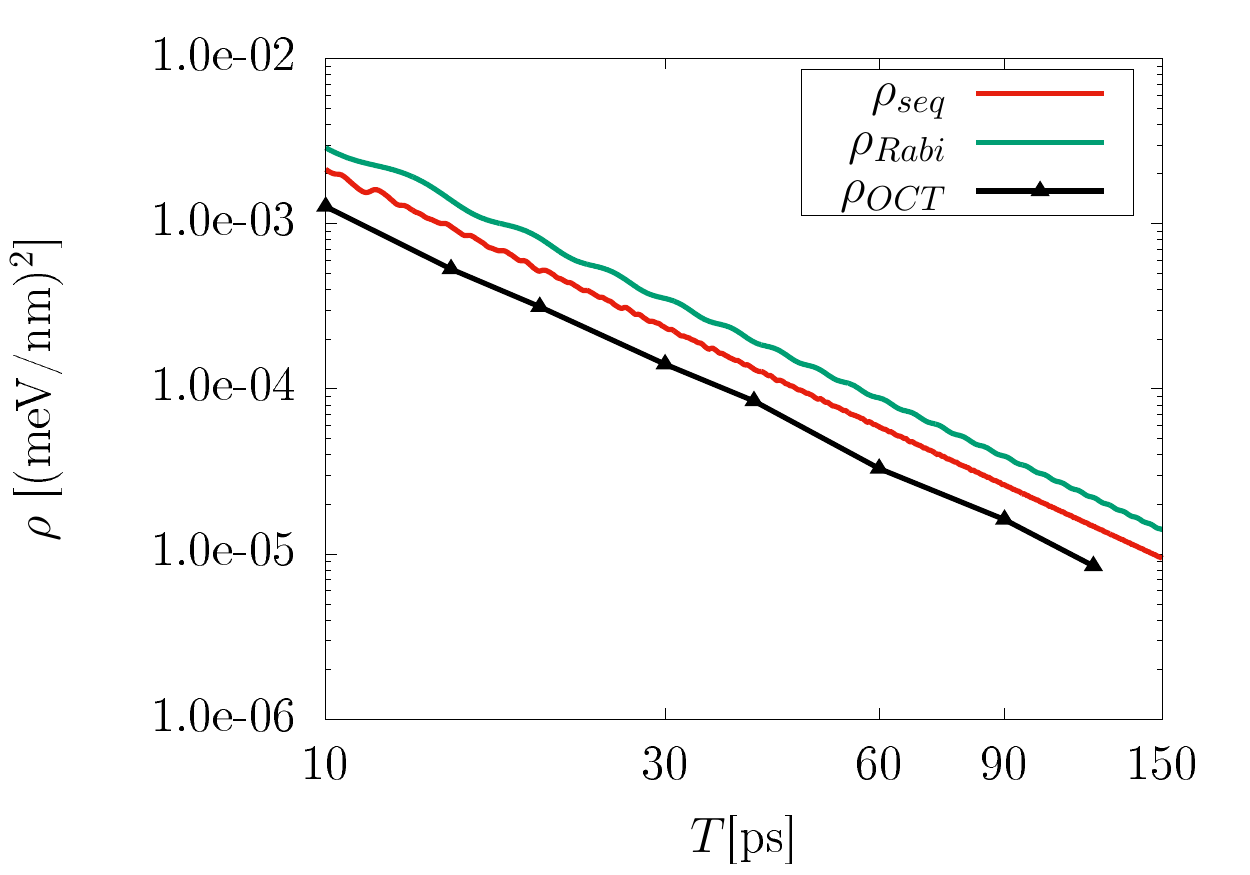} 
\end{array}$
\caption{\label{fluence-T} Mean square amplitude of the field for a several OCT designed pulses having 10 ps $\leq T \leq$ 150 ps $\rho_{OCT}$, as compared with two- (Rabi)  $\rho_{Rabi}$ and three-level based models $\rho_{seq}$ (see text for further details) }
\end{figure}

The relation between the fluence and length of the pulses for high yield transitions $|0^{1/2}\rangle\rightarrow|1^{1/2}\rangle$ required for quantum information processing was further investigated. Fig. \ref{fluence-T} shows the locus of high fidelity processes ($\sim 0.9999$) in the plane time-fluence for direct Rabi transitions, OCT designed pulses and sequential pulses. As it can be seen, all of them satisfy an approximate power-law $\rho\sim T^{-\gamma}$ with $\gamma\approx 2$.

 The known RWA solution for the two-level process, provides an analytic relation between the fluence and the time length $T$ for a monochromatic Rabi pulse of frequency $\omega_{10}$ transferring completely the population from the initial to the target state, giving
\begin{eqnarray}\label{rho2L}
\rho_{\rm Rabi} =\left(\frac{\pi}{\mu_{01}}\right)^2\frac{ (2 T \omega_{01}+\sin (2 T \omega_{01}))}{4 \omega_{01}T^3}.
\end{eqnarray}
For times $T$ much longer than $\omega_{10}^{-1}$, the second term can be neglected and $\rho\sim T^{-2}$.

We could perform the same analysis for the indirect strategy $0 \rightarrow 2 \rightarrow 3 \rightarrow 1 $. 

The conditions for the dynamics of the sequential pulse (\ref{pulse seq}) within the RWA, as given in  \ref{ApendixA}, are as follows: the ground state is initially occupied, and at some time $\tau$ its population is completely transferred to the second and third excited states. Then these two states are depopulated, and the probability is transferred to the target initially depopulated. In \ref{ApendixA} it is shown that $\tau=T/2$.  
The the mean square amplitude for this model is given by 
\begin{equation}
\rho_{seq}=\frac{1}{T} \int_{0}^{T} F_{seq}(t)^2 dt    
\label{rho_seq}
\end{equation}
In Fig. \ref{fluence-T} we compare the mean square amplitude as a function of the control time $T$ for the Rabi oscillation (\ref{rho2L}), the sequential strategy (\ref{rho_seq}), and the OCT results. The OCT results correspond to the lowest fluence in which the target can be reached with a yield of 99.99\%. For any given $T$, the fluence for the Rabi strategy is always larger than the sequential pulse and the OCT protocol, {\em i.e.}, $\rho_{\rm Rabi}> \rho_{\rm seq} > \rho_{\rm OCT}$. 
In the Figures \ref{OCT} (c) and (d)  we show the time evolution of the state probabilities for $T$=60 and 90 ps obtained with OCT (solid line) and the sequential pulse model (dashed line). Similarity of both strategies shows that the mechanism involved in the OCT dynamics are basically described by $F_{\rm seq}(t)$. The differences could be reduced by including envelope profiles in the model.

\section{Conclusions}
In this work, we aimed to design fast pulses for accurate controlling a qubit encoded in the three-electron states of a triple quantum dot.

We have shown that driving fields in the optical range can be designed by applying optimal control theory, while still satisfying  prescribed sub-nanosecond operation times, experimentally available field amplitudes and high fidelity transfer to the target state.

The pulses obtained without any frequency cutoff, have transfer times between the qubit states of the order of tens to hundreds of picoseconds with the fidelity required for quantum information processing. Their frequency spectra correspond, mainly, to the energy gaps between the qubit states as well as between some of the closer excited states and the qubit states.

If the frequencies present in the pulse spectrum are restricted to be under the THz range, as in experiments with gate voltage control, only direct Rabi transitions between the qubit states are found. On the other hand, by allowing optical range frequencies to be present in the designed pulse, fast switching process involving intermediate transitions to the closer excited states are obtained.

Since the dynamics of the TQD strongly depends on the precise representation of both the electronic structure and the transition dipole moments, we compared our configuration-interaction calculations to the usual description provided by the extended Hubbard model. 
Our results show that, although a proper parametrization of the Hubbard model, can provide reasonable agreement to the CI calculated level structure, it cannot do so for the electric dipole transition moments as it is needed for the process considered, so that it is not reliable for designing control strategies involving transitions to excited states such as the proposed in this work.

Finally, our assumption that the external electric field is uniform is unnecessary, although the net effect of the space modulation of the electric field would be to reduce the switching rate or to give rise to a more complicated pulse sequence. 

\section*{Aknowledgements}

We acknowledge SECYT-UNC, SGCyT-UNNE (PI 17F020) and CONICET (PIP-11220150100327CO, PUE2017-22920170100089CO, PIP-11220130100361CO) for partial financial support. O.O. wishes to acknowledge the warm hospitality of the Nanophysics Group at FACENA and IMIT-CONICET, Corrientes, where this work was done.  

\begin{appendix}
\section{Effective six-level Hubbard model\label{Apendix-hubbard}}
By chosing the basis $\{| n\rangle\}$ ($n=1,\ldots, 6$) \cite{revbuk}, of single occupied ($|0\rangle$ and $|1\rangle$) and double occupied configurations ($|2\rangle$, $|3\rangle$, $|4\rangle$ and $|5\rangle$),
\begin{eqnarray}
| 0 \rangle &=& \frac{1}{\sqrt{2}} \left(- | \Phi_{123}\rangle + | \Phi_{231}\rangle\right) \\
| 1 \rangle &=& \frac{1}{\sqrt{6}} \left(  | \Phi_{123}\rangle + | \Phi_{231}\rangle\right) -\sqrt{\frac{2}{3}}| \Phi_{312}\rangle \\
| 2 \rangle &=&    | \Phi_{311}\rangle \\
| 3 \rangle &=&    | \Phi_{133}\rangle \\
| 4 \rangle &=&    | \Phi_{211}\rangle \\
| 5 \rangle &=&    | \Phi_{233}\rangle, 
\label{basis}
\end{eqnarray}
the effective 6-level Hubbard approximation takes the form
\begin{widetext}
\begin{equation}
H_{\rm Hub} = 
\left(
\begin{array}{cccccc}
 0 & 0 & \frac{\tau }{2} & \frac{\tau }{2} & \frac{\tau }{2} & \frac{\tau
   }{2} \\
 0 & 0 & \frac{\sqrt{3} \tau }{2} & -\frac{\sqrt{3} \tau }{2} &
   -\frac{\sqrt{3} \tau }{2} & \frac{\sqrt{3} \tau }{2} \\
 \frac{\tau }{2} & \frac{\sqrt{3} \tau }{2} & V _1-V _2+U-2
   U_C & 0 & 0 & 0 \\
 \frac{\tau }{2} & -\frac{\sqrt{3} \tau }{2} & 0 & -V _2+V
   _3+U-2 U_C & 0 & 0 \\
 \frac{\tau }{2} & -\frac{\sqrt{3} \tau }{2} & 0 & 0 & V _2-V
   _3+U & 0 \\
 \frac{\tau }{2} & \frac{\sqrt{3} \tau }{2} & 0 & 0 & 0 & -V _1+V
   _2+U
\end{array}
\right).
\end{equation}
\end{widetext}
The dipole moment in the 3-particle basis $| n\rangle$ becomes
\begin{widetext}
\begin{equation}
\mu
=  
\left(
\begin{array}{cccccc}
 0 & 0 & -\frac{\nu }{\sqrt{2}} & \frac{\nu }{\sqrt{2}} & \frac{\nu }{\sqrt{2}} & -\frac{\nu }{\sqrt{2}} \\
 0 & 0 & -\sqrt{\frac{3}{2}} \nu  & -\sqrt{\frac{3}{2}} \nu  & -\sqrt{\frac{3}{2}} \nu  & -\sqrt{\frac{3}{2}} \nu  \\
 -\frac{\nu }{\sqrt{2}} & -\sqrt{\frac{3}{2}} \nu  & \lambda  & 0 & 0 & 0 \\
 \frac{\nu }{\sqrt{2}} & -\sqrt{\frac{3}{2}} \nu  & 0 & -\lambda  & 0 & 0 \\
 \frac{\nu }{\sqrt{2}} & -\sqrt{\frac{3}{2}} \nu  & 0 & 0 & \lambda  & 0 \\
 -\frac{\nu }{\sqrt{2}} & -\sqrt{\frac{3}{2}} \nu  & 0 & 0 & 0 & -\lambda  \\
\end{array}
\right)
\end{equation}
\end{widetext}

Assuming  $\lambda=\langle\varphi_R|e x|\varphi_R\rangle=-\langle\varphi_L|e x|\varphi_L\rangle$ is the diagonal matrix elements of the dipole moment when evaluated with functions localized at the left (right) dots, and $\nu=\langle\varphi_L|e x|\varphi_C\rangle\sim \mu S_{LC} \ll \lambda$, $S_{LC}$ being the overlap between basis functions located at the left (or right) and center dots. In the approximate model of the LTQD, $\lambda\approx 1$ and $\nu\approx 0$.
\section{RWA solution of Schr\"odinger equation for the sequential pulse $F_{\rm seq}(t)$ \label{ApendixA}}
We present here the time-dependent solution for the dynamics of the LTQD states driven by the pulse $F_{\rm seq}(t)$, eq. (\ref{pulse seq}). 

Defining $\Omega_{02}=\mu_{02} A$ and $\Omega_{23}=\mu_{23} B$ and neglecting rapidly oscillating terms (RWA), the Schr\"odinger equation in the interaction picture, reads

{\em Ladder-process: } $\omega_a=\omega_{20}$, $\omega_b=\omega_{32}$
\begin{eqnarray}
i\dot{c}_0(t) &=&\frac{\omega_{02}}{2} c_{2}  \\
i\dot{c}_2(t) &=& \frac{\omega_{02}}{2} c_0 + \frac{\omega_{23}}{2} c_{3}  \\
i\dot{c}_3(t) &=& \frac{\omega_{23}}{2} c_{2}   
\end{eqnarray}

$\Lambda$-{\em process: } $\omega_a=\omega_{31}$, $\omega_b=\omega_{32}$
\begin{eqnarray}
i\dot{c}_{2}(t) &=&\frac{\Omega_{23}}{2} c_{3} \nonumber \\
i\dot{c}_{3}(t) &=& \frac{\Omega_{23}}{2} c_{2} + \frac{\Omega_{02}}{2} c_{1}  \\
i\dot{c}_{1}(t) &=& \frac{\Omega_{02}}{2} c_{3}   \nonumber
\end{eqnarray}

By imposing the initial and final conditions 
\begin{eqnarray}
c_0(0)&=&1, \hspace{0.5cm}  c_{2}(0)=c_{3}(0)=0, \\
c_1(T)&=&1, \hspace{0.5cm} c_{2}(T)=c_{3}(T)=0,
\end{eqnarray}
the following solutions are obtained

{\em Ladder-process: }

\begin{eqnarray}\label{three_a}
c_0(t) &=& \left( \frac{\Omega_{02}}{\Delta} \right)^2 \cos{\left(\frac{\Delta}{2} t \right)} + \left(\frac{\Omega_{23}}{\Delta}\right)^2  \nonumber \\
c_{2}(t) &=& -i \left(\frac{\Omega_{23}}{\Delta}\right) \sin{\left(\frac{\Delta}{2} t \right)}  \\
c_{3}(t) &=& \left(\frac{\Omega_{02} \Omega_{23}}{\Delta^2}\right) \left( \cos{\left(\frac{\Delta}{2} t \right) } -1 \right)  \nonumber
\end{eqnarray}

$\Lambda$-{\em process: }

\begin{eqnarray}\label{three_b}
c_{1}(t) &=& \left( \frac{\Omega_{02}}{\Delta} \right)^2 \cos{\left(\frac{\Delta}{2} (t-T) \right)} + \left(\frac{\Omega_{23}}{\Delta}\right)^2  \nonumber \\
c_{3}(t) &=& -i \left(\frac{\Omega_{23}}{\Delta}\right) \sin{\left(\frac{\Delta}{2} (t-T) \right)}  \\
c_{2}(t) &=& \left(\frac{\Omega_{02} \Omega_{23}}{\Delta^2}\right) \left( \cos{\left(\frac{\Delta}{2} (t-T) \right) } -1 \right)  \nonumber
\end{eqnarray}

Taking $\tau=T/2$ the pulse becomes symmetric, as observed in the OCT pulses, and results in a complete depletion of both the initial and target states at $T/2$:  $|c_0(T/2)|^2=0=|c_1(T/2)|^2$. The continuity of the solutions at $T/2$ also implies $ \Omega_{02} = \sqrt{2} \Omega_{23}$ and 
\begin{equation}
T= \frac{4}{\Delta}\arccos{\left(-\frac{\Omega_{23}^2}{\Omega_{02}^2} \right),}
\end{equation}
therefore, $T= 8\pi /3\sqrt{3} \Omega_{23}$.

\end{appendix}

\end{document}